\documentstyle[prb,aps,graphicx]{revtex}  
\def\ET{BEDT-TTF}               
\def\ETI{(\ET)$_2$I$_3$}        
\def\AETI{$\alpha$-\ETI}        
\def\BETI{$\beta$-\ETI}         
\def\KETI{$\kappa$-\ETI}        
\def\k{{\bf k}}                 
\def\q{{\bf q}}                 
\begin{document}
\draft
\title{Lattice dynamics and electron-phonon coupling in
{\BETI} organic superconductor}

\author{Alberto Girlando, Matteo Masino, and Giovanni Visentini}

\address{Dipartimento di Chimica Generale ed Inorganica, Chimica Analitica,
Chimica Fisica, Universit\`a di Parma,
Parco Area delle Scienze, I-43100, Parma, Italy}

\author{Raffaele Guido Della Valle, Aldo Brillante, and Elisabetta Venuti}

\address{Dipartimento di Chimica Fisica e Inorganica,
Universit\`a di Bologna, Viale Risorgimento 4, I-40136 Bologna, Italy}


\date{\today}
\maketitle

\begin{abstract}

The crystal structure and lattice phonons of
{\ETI} superconducting $\beta$-phase
(where  {\ET} is bis-ethylen-dithio-tetrathiafulvalene)
are computed and analyzed by the Quasi Harmonic Lattice
Dynamics (QHLD) method. The empirical atom-atom potential is that successfully
employed for neutral {\ET} and for non superconducting {\AETI}.
Whereas the crystal structure and its temperature
and pressure dependence are properly reproduced within a rigid molecule
approximation, this has to be removed
account for the specific heat data. Such a 
mixing between lattice and low-frequency {\it intra}--molecular vibrations
also yields good agreement with the observed
Raman and infrared frequencies. From the
eigenvectors of the low-frequency phonons we calculate
the electron-phonon coupling constants due to the modulation
of charge transfer (hopping) integrals. The charge transfer integrals
are evaluated by the extended H\"uckel method applied to all nearest-neighbor
{\ET} pairs in the $ab$ crystal plane. From the averaged electron-phonon
coupling constants and the QHLD phonon density of states we
derive the Eliashberg coupling function $\alpha(\omega)F(\omega)$,
which compares well with that experimentally obtained
from point contact spectroscopy. The corresponding
dimensionless coupling constant $\lambda$ is found to be 
$\sim 0.4$.

\end{abstract}


\section{Introduction}
The strength and importance of electron-lattice phonon ({\it
e--lph}) coupling in the superconductivity mechanism of organic
superconductors has always been rather controversial. Early numerical
estimates based on simplified models gave very low values for the coupling to
acoustic phonons,\cite{Berlinsky1974} and much more
attention was then devoted to electron-molecular vibration {\it e--mv}
coupling. \cite{Yamaji1987} On the other hand, ``librons'' were also invoked in the pairing mechanism of
organic superconductors.\cite{Gutfreund1980,Nowack1986} 
On the experimental side, most of the data have been collected for
bis-ethylen-dithio-tetrathiafulvalene ({\ET}) salts, which are the most
extensive and representative class of organic superconductors.\cite{Ishiguro} In
particular, recent Raman experiments on {\ET} salts pointed out that the
intensity \cite{Pokhodnia1993} and the frequency \cite{Pedron1997} of
some low-frequency phonon mode change at the superconducting
critical temperature $T_c$. Oddly enough, also one {\it intra}--molecular
{\ET} mode has been shown to exhibit a frequency shift at $T_c$.
\cite{Eldridge1998} Carbon isotopic substitution on the central double
bond {\ET} was claimed to have dramatic effects on the $T_c$
of one superconducting {\ET} salt,\cite{Merzhanov1992} but subsequent extensive
isotopic substitution studies on other superconducting {\ET} salts strongly
suggested that the lattice phonons are likely involved in the superconducting
mechanism.\cite{Kini1996} Attempts to take into account both {\it e-mv} and
{\it e-lph} coupling have been put forward,\cite{Pedron1993} but the role
and the relative importance of the two types of coupling in the pairing
mechanism is far from being settled.

Whereas extensive studies have been devoted to the characterization of
{\it intra}--molecular phonons of {\ET} \cite{Eldridge1995,Demiralp1997,Liu1997}
and to the estimate of the relevant {\it e--mv} coupling strength,\cite{Visentini1998}
very little is known about the lattice phonon structure in {\ET} salts
or in other organic superconductors. Obtaining a sound
characterization of {\ET} salts lattice phonons is not easy, since in
general the unit cell contains several molecular units, and the phonon
modes obviously differ for different crystalline structures. We have
tackled the problem by adopting the ``Quasi Harmonic Lattice
Dynamics'' (QHLD) method,\cite{Ludwig1967,Valle1995,Valle1996} by which we are able
to analyze both the crystal and the lattice phonon structure in terms
of empirical atom-atom potentials, in principle transferable among
crystals containing the same atoms. We have first obtained C, S and H
atom-atom potential parameters reproducing crystal structure and lattice phonons
of neutral {\ET}.\cite{Brillante1997} Then we have considered the I$_3^-$
salts, which have only one additional atom to parametrize, and present
several crystalline phases.\cite{Ishiguro} After the successful
application of the potential to non-superconducting {\AETI} crystal,
\cite{Valle1999} we present in this paper the results relevant to the
extensively studied superconducting $\beta$-phases.

{\BETI} has been the first ambient pressure {\ET} based superconductor
to be discovered,\cite{Yagubskii1984} and its unit cell
contains only one formula unit.\cite{Mori1984} The {\ET} radicals are
arranged in stacks, and the stacks  form sheets
parallel to the {\it ab} crystal plane. The centrosymmetric
linear I$_3^-$ anions separate the sheets, forming an insulating layer.
Several variants of the {\BETI} phase have been reported,
making difficult a full and detailed characterization.
The electrochemically prepared {\BETI} exhibits ambient pressure
superconductivity at $T_c$= 1.3 K ($\beta_L$-{\ETI}) or at $T_c$=8.1
($\beta_H$- or $\beta^*$-{\ETI}) depending on the pressure-temperature
history of the sample. Such $T_c$ increase
has been attributed to a pressure induced ordering process of the
ethylene groups of {\ET} cation.\cite{Schultz1986}
In addition, thermal treatment or laser
irradiation of the $\alpha$-phase yields an irreversible transformation to a
superconducting phase ($T_c$= 8.0 K), named $\alpha_t$-{\ETI},\cite{Pokhodnia1993} which was
claimed to be similar to the $\beta$-phase. On the other hand, {\BETI}
can also be prepared by direct chemical oxidation ($\beta_{CO}$-{\ETI}),\cite{Mueller1997}
with $T_c$ between 7.1 and 7.8 K. Recent X-ray data confirm that
thermally treated {\AETI} is identical to $\beta_{CO}$-{\ETI},\cite{Mueller1999}
but it is still not clear whether $\beta_{CO}$-{\ETI} is the same as
$\beta_H$-{\ETI}: the possibility of non-stoichiometric phases has also
been put forward as an alternative to the ordering process in causing
a $T_c$ of about 8 K.\cite{Madsen1999}

The paper is organized as follows. We first discuss in some detail
the methods we have adopted to calculate the structure, the
phonon dynamics and the {\it e--lph} coupling strength of {\ETI}
salts. The results relevant to the {\BETI} phase are then presented and
compared with available experimental data. Finally, the possible role of
electron-phonon coupling in the pairing mechanism of organic superconductors
is briefly discussed.

\section{Methods}
\label{sec:methods}

\subsection{Quasi Harmonic Lattice Dynamics}

The crystal structure at thermodynamic equilibrium of {\ETI} salts
is computed using Quasi Harmonic Lattice Dynamics (QHLD).
In QHLD \cite{Ludwig1967,Valle1995,Valle1996} the Gibbs free energy $G(p,T)$ of the crystal
is approximated with the free energy of the harmonic phonons calculated at
the average lattice structure ($\hbar = 1$):

$$ G(p,T) = \Phi_{\rm inter} + pV+ \sum_{\q i} \frac{\omega_{\q i}}{2}
+ k_B T ~ \sum_{\q i} \ln\left[1 - \exp\left(-\frac{\omega_{\q i}}{k_BT}\right)
\right] \eqno(1) $$

\noindent
Here, $\Phi_{\rm inter}$ is the total potential energy of the crystal, $pV$ is
the pressure-volume term, $\sum_{\q i} \omega_{\q i}/2$ is the zero-point
energy, and the last term is the entropic contribution. The sums are
extended to all phonon modes of wavevector {\q} and
frequency $\omega_{\q i}$. Given an initial lattice structure,
one computes $\Phi_{\rm inter}$ and its second derivatives with respect to
the displacements of the molecular coordinates. The second derivatives
form the dynamical matrix, which is numerically diagonalized to obtain
the phonon frequencies $\omega_{\q i}$ and the corresponding
eigenvectors. The structure as a function of $p$ and $T$ is then
determined self-consistently by minimizing $G(p,T)$ with respect to
lattice parameters, molecular positions and orientations.

In the case of {\ETI} salts, and in particular of the $\beta$-- phase,
the choice of the initial lattice structure is somewhat problematic,
due to the conformational disorder of the {\ET} molecules.  In fact,
the X-rays structural investigations \cite{Leung1985} indicate that {\BETI}
at 120 K is disordered with two alternative sites for the terminal C
atoms, labeled 9a,10a (staggered form) and 9b,10b (eclipsed form).
On the other hand, {\it ab--initio} calculations \cite{Demiralp1995} for neutral {\ET} indicate that the ``boat'' geometry
($C_2$ symmetry) is more stable than the ``planar'' geometry ($D_2$
symmetry) by 0.65 kcal/mole. The ``chair'' distortion ($C_s$ symmetry)
is slightly more stable than the planar molecule, but still less
stable than the boat one. The {\ET}$^+$ ion is planar, and in {\ETI}
crystals we have a statistical mixture of neutral and ionized
molecules. On the basis of the site symmetry constraints, we observe
that neutral molecule boat and chair geometries correspond to
the Leung's configurations 9a,10a and 9b,10b, respectively.\cite{Leung1985}
Thus the conformational disorder observed in most {\ET} salts is readily
understood: the energetic cost of deforming the molecules is small
with respect to the energy gain among different packing
arrangements in the crystals. To investigate at least partially the
effect of conformational disorder on the stability of {\ETI} phases,
we have performed several calculations starting from different
initials molecular geometries, as detailed in Section III.

\subsection{Potential Model}

We have adopted a pairwise additive inter-molecular potential of the form
$\Phi_{\rm
inter}=\frac{1}{2}\sum_{mn} [q_m q_n/r_{mn} +
A_{mn}\exp(-B_{mn}r_{mn}) - C_{mn}/r_{mn}^6] $
where the sum is
extended to all distances $r_{mn}$ between pairs $m$,$n$ of atoms in
different molecules. The Ewald's method \cite{Califano1981} is used
to accelerate the convergence of the Coulombic interactions $q_m q_n /
r_{mn}$. The atomic charges $q_m$ are the PDQ (PS-GVB) results of a recent
{\it ab--initio} Hartree-Fock calculations, \cite{Demiralp1994} and
are introduced to model both the neutral and ionized forms of the
{\ET} molecule. The parameters $A_{mn}$, $B_{mn}$ and $C_{mn}$
involving C, H and S atoms are taken from our previous
calculation of neutral {\ET}.\cite{Brillante1997} Since in the chosen
model \cite{Hall1975} C--H parameters
are computed from C--C and H--H parameters via ``mixing rules'',
the same procedure is adopted here for all the interactions
between different types of atoms.
The iodine parameters have been derived from 9,10-diiodoanthracene,
and successfully tested on {\AETI}.\cite{Valle1999} The complete atom-atom
model is given in Table I.

\subsection{Specific Heat}

The constant volume specific heat as a function of $T$ is computed
directly from its statistical mechanics expression for a
system of phonons:

\def\arg{\frac{\omega_{\q i}}{k_B T}}

$$ C_V(T) = \sum_{\q i} k_B \left(\arg\right)^2 \exp\left(-\arg\right)
 \left[1-\exp\left(-\arg\right)\right]^{-2} \eqno(2) $$

\noindent
As usual in these
cases, eq. (2) is evaluated by sampling a large number of $\q$-vectors
in the first Brillouin Zone (BZ).

In our first attempts to compute $C_V$, we sampled over regular grids
in the BZ. We have found that for $T\le5$ K the statistical noise was
still noticeable even after summing over several thousands of
$\q$-vectors; the results were dependent on the sample size. At
large $T$, on the contrary, statistical convergence was quite fast.
This pathology can be attributed to the fact that, due to the
exponential factor in eq. (2), only the phonons with $\omega_{\q i} \le
k_B T$ give a non-negligible contribution to $C_V(T)$. For very low
$T$, only the acoustic branches of the phonons with $\q$ close to
zero have sufficiently small frequencies. With a regular grid, only a
few of these vectors are sampled, and most of the computer time is
wasted over regions of the BZ that are already well sampled.

To obtain accurate statistics at a reasonable cost, we have used a
Monte Carlo (random) integration scheme, biased to yield a larger
sampling probability close to $\q=0$. For computational simplicity,
we have chosen a three-dimensional Lorentzian probability
distribution, $L(\q)\propto(1+a\mid\q\mid^2)^{-1}$, where $a$ is a
width parameter. The bias is compensated by using the reciprocal of
the sampling probability as the sample weight. With this scheme most
of the computer effort is spent in the region $\q\approx0$, where a
denser sampling really matters. By summing over about 2000
$\q$-vectors, we have been able to reach a satisfactory statistical
convergence, in the whole range between 0.1 and 20 K. At high $T$,
the results coincide with those obtained by integrating over a grid.
At low $T$, $C_V$ goes as $T^3$, as it should when the acoustic modes
are properly sampled, and does not fluctuate with the sample size.

\subsection{Coupling with low-frequency {\it intra}--molecular
degrees of freedom}

In most calculations for molecular crystals all {\it intra}--molecular
degrees of freedom are neglected and the molecules are maintained
as rigid units. This rigid molecule approximation (RMA) is reasonable
for small compact molecules, like benzene, where all normal
modes have frequencies much higher than those of the lattice phonons.

Since for both I$_3^-$ and {\ET} several investigations \cite{Demiralp1997,Liu1997,Dressel1992} suggest that there are low frequency
{\it intra}--molecular modes, the validity of RMA for {\ETI} appears
questionable. Therefore, we have decided to relax the RMA and to
investigate the effects of the {\it intra}--molecular degrees of freedom.
For this purpose we adopt an exciton-like model.\cite{Califano1981} 
To start with, it is convenient to use a set of
molecular coordinates $Q_i$ describing translations, rotations and
internal vibrations of the molecular units in the crystal.
To each {\ET} molecule of $N=26$
atoms we associate the following $3N$ coordinates: 3 mass-weighted
cartesian displacements of the center of mass, 3 inertia-weighted
rotations about the principal axes of inertia, and $3N-6=72$ internal
vibrations (the normal modes of the isolated {\ET} molecule). The
I$_3^-$ ion, which is linear, has 3 translations, 2 rotations and $4$ internal
vibrations. In order to compute the phonon frequencies, we need
all derivatives $\partial^2 \Phi/\partial Q_{ri} \partial Q_{sj}$
of the total potential $\Phi$ with respect to all pairs of molecular
coordinates $Q_{ri}$ and $Q_{sj}$. Here $r$ and $s$ label molecules
in the crystal, while $i$ and $j$ distinguish molecular coordinates.

The potential $\Phi$ is made of {\it intra}-- and {\it inter}--molecular parts,
$\Phi_{\rm intra}$ and $\Phi_{\rm inter}$. In the exciton model, the
diagonal derivatives of $\Phi_{\rm intra}$ potential are taken to
coincide with those of an isolated molecule: $\partial^2 \Phi_{\rm
intra}/\partial Q_{ri}^2=\omega^2_{ri}$. Here $\omega_{ri}$ is the
frequency of the $i$-th normal mode of the $r$-th molecule. All
off-diagonal derivatives are zero, which means no coupling among
different normal modes, and no coupling between normal modes and rigid
roto-translations. These assumptions are correct for the
{\it intra}--molecular potential at the harmonic level (by definition).

The coupling between the molecular coordinates is given by
$\Phi_{\rm inter}$. For {\BETI}, $\Phi_{\rm inter}$ is
described by atom-atom and charge-charge interactions, which are both
functions only of the interatomic distance. Since the distance
depends on the cartesian coordinates of the atoms, $X_{ra}$, the
derivatives of $\Phi_{\rm inter}$ can be directly computed in terms
of the coordinates $X_{ra}$, and then converted to molecular
coordinates $Q_{ri}$:

$$
\frac{\partial^2 \Phi_{\rm inter}}{\partial Q_{ri} \partial Q_{sj}} =
\sum_{ab} \frac{\partial^2 \Phi_{\rm inter}}{\partial X_{ra}
   \partial X_{sb}} ~~\frac{\partial X_{ra}}{\partial Q_{ri}} ~~
   \frac{\partial X_{sb}}{\partial Q_{sj}} \eqno(3) $$

\noindent
Here $a$ and $b$ label the cartesian coordinates of the atoms in
molecules $r$ and $s$, respectively, and the matrix $\partial
X_{pa}/\partial Q_{pi}$ describes the cartesian displacements which
correspond to each molecular coordinate $Q_{pi}$. The displacements
corresponding to rigid translations and rotations of the molecules can
be derived by simple geometric arguments. The
displacements associated to the {\it intra}--molecular degrees of freedom are
the cartesian eigenvectors of the normal modes of the isolated
molecule. The atomic displacements, together with
the {\it inter}--molecular potential model, determine the coupling between
{\it intra}--molecular and lattice modes. We remark that the
{\it intra}--molecular degrees of freedom are
taken into account only as far as their effects on the vibrational
contribution to the free energy are concerned. No attempt to decrease
the potential energy by deforming the molecules is done.

\subsection{{\it e--lph} coupling constants and the Eliashberg function}

In molecular crystals, {\it intra}--molecular vibrations are assumed
to couple with electrons through modulation of on--site energies
({\it e--mv} coupling). Lattice phonons are instead expected to
modulate mainly the {\it inter}--molecular charge transfer (CT)
integral, $t$, the corresponding linear {\it e--lph} coupling constants
being defined as:

$$g(KL;\q,j) = (\partial t_{KL}/{\partial Q_{\q j}}) \eqno(4)$$

\noindent
where $t_{KL}$ is the CT integral between neighboring pairs K,L of {\ET}
molecules, and 
where $Q_{\q j}$ is the dimensionless normal coordinate for
the $j$--th phonon with wavevector $\q$.
By relaxing the RMA, as explained above,
the distinction between low--frequency {\it intra}--molecular
modes and lattice modes is at least partially lost.
On the other hand, {\it e--mv} coupling by the low-frequency
molecular modes is expected to be fairly small, as suggested
by the calculations available for isolated
{\ET}.\cite{Demiralp1997,Liu1997} Therefore,
we have assumed that the calculated low--frequency phonons of {\BETI},
occurring between 0 and about 200 cm$^{-1}$, are coupled to the
CT electrons only through the $t$ modulation.

To evaluate the $g(KL;\q,j)$'s, we have followed a real space
approach.  Adopting the extended H\"uckel method, for each pair $K,L$ of
{\ET} molecules within the {\BETI} crystal we have calculated
$t_{KL}$  as the variation of the HOMO energy in going from the monomer
to the dimer. Such an approach is known to give $t$ values
in nice agreement with those calculated by extended basis
set {\it ab--initio} methods.\cite{Fortunelli1997} 
$t_{KL}$ is calculated for the dimer equilibrium geometry 
within the crystal, as
well as for geometries displaced along the QHLD eigenvectors.
The various $g(KL;\q,j)$ are then obtained by numerical differentiation.
We have considered only the modulation of the four largest
$t$'s, all along the $ab$ crystal plane.

In the case of {\it e--mv} coupling the overall electron-phonon coupling
strength is generally expressed by the small polaron binding energy,
$E_{sp}^{mv} = \sum_{i} g_i^2 / \omega_{i}$,
where both $g_i$, the $i$-th {\it e--mv} coupling constant,
and $\omega_{i}$, the corresponding
reference frequency, are quite naturally taken as independent of
the wavevector $\q$.\cite{Visentini1998}  Also in the calculation 
the {\it e--lph} coupling we have assumed the optical
lattice phonons as dispersionless, and have performed
the calculations for the $\q =0$ eigenvectors only. Within this approximation,
symmetry arguments show that only the totally symmetric ($A_g$) phonons
can be coupled with electrons. Thus, the overall {\it e--lph}
coupling strength for the $j$-th 
lattice {\it optical} phonon, can again be expressed
by the small polaron binding energy relevant to the $j$-th phonon:
$\epsilon_j =\sum_{KL}(g_{KL,j}^2/\omega_j)$.
The total coupling strength is then given by $E_{sp}^{lp} = \sum_j \epsilon_j$.

For the three acoustic branches we must of course consider the
$\q$ dependence of the $g$'s, the coupling constants being zero
for $\q$ = 0. We have then calculated the coupling strength
($\epsilon_j^{ac}$) at some representative BZ edges 
in the $a^*b^*$ reciprocal plane. For each branch,
we have averaged the found $\epsilon_j^{ac}$, and assumed 
a linear dependence on $|\q|$. The latter
assumption is correct only in the small $|\q|$ limit.\cite{Allen}
 
The most important single parameter characterizing the strength of
electron-phonon coupling in the superconductivity mechanism is
the dimensionless electron--phonon coupling constant
$\lambda$.\cite{Allen,Parks} This parameter is in turn related
to the Eliashberg coupling function $\alpha^2(\omega)F(\omega)$:\cite{Parks}

$$ \lambda = 2 \int_0^{\omega_{max}} \frac{\alpha^2(\omega)
F(\omega)}{\omega}\,\rm d \omega \eqno(5)$$
where $F(\omega)$ is the phonon density of states per unit cell,
and $\alpha^2(\omega)$
is an effective coupling function for phonons of energy $\omega$.
The {\it e-lph} Eliashberg coupling function can be evaluated from
the QHLD phonon density of states and from the electron--phonon
matrix element $\sl{g}(\bf{k,k'}\rm ;j)$ expressed in the
reciprocal space:\cite{Allen}

$$\alpha^2(\omega)F(\omega) = N(E_F)\sum_{\j}\langle|{\sl g}(\k,\k';j)|^2
\delta(\omega-\omega_{\q j})\rangle_{FS} \eqno(6)$$
where ${\q} = {\k'} - {\k}$, $\k$ and $\k'$ denoting the electronic
wavevectors, and $N(E_F)$ is the density of states per spin per
unit cell at the Fermi level. In eq.(6), $\langle\,\, \rangle _{FS}$
indicates the average over the Fermi surface.

We have calculated the $g$'s in real space, as detailed above.
In order to introduce the dependence on the electronic wavevector $\k$,
as required in eq. (6), we have to describe the electronic structure of
the $\beta$--phase metal. To get a simple yet realistic model we make
resort of the rectangular tight--binding dimer model,\cite{VisentiniK}
where the {\ET} dimers inside the actual
unit cell are taken as a supermolecule. Actually, as in $\kappa$--phase,
in the $\beta$--phase structure {\ET} dimers are clearly recognized
(in the present formalism, they correspond to the $t_{AB}$ CT integral).
In this model there is only one half--filled conduction band in the
first BZ, whose dispersion relation as a function of the $t_{KL}$ CT
integrals is easily obtained:

$$\epsilon(\k) = t_{AB} + t_{AH}~\cos(k_x) + t_{AE}~\cos(k_y) +
 t_{AC}~\cos(k_x+k_y) \eqno(7)$$

The chemical potential is obtained numerically from the half--filling
condition. Within our tight--binding approximation, the dependence
in reciprocal space of the coupling constants associated to the
{\it inter}--dimer (inter--cell) hoppings is given by:\cite{Conwell1980}

$${\sl g}(\k,\k';j) = 2 {\rm i}~g(KL;\q,j) [\sin({\bf k}+{\bf q}){\bf R}
- \sin{\bf k}{\bf R}] \eqno(8)$$
\noindent
where {\bf R} represents the nearest--neighbor lattice vectors
($a$, $b$, $a+b$), and $g(KL;\q,j)$ are the three corresponding
real space inter--cell CT integrals. The Fermi surface average of eq. (6)
can now easily performed numerically for the inter--dimer contribution.
The coupling constants associated with the modulation of the
{it intra}--dimer CT integrals are treated as intramolecular
coupling constants, and as such are independent
of $\k$.\cite{Pedron1993,Conwell1980}
We finally remark that the {\it e-mv} Eliashberg coupling
function is simply given by $[\alpha^2(\omega)F(\omega)]_{e-mv} =
(N(E_F)/N)\sum_i g^2_i \delta (\omega - \omega_i)$, $N$ being the
number of molecules per unit cell and $g_i$ being the usual
{\it e-mv} coupling constant,\cite{Pedron1993} so that $\lambda_{e-mv} =
N(E_F) E_{sp}^{mv}$.

\section{Results}
\label{sec:results}

\subsection{Crystallographic structures}

The unit cell of {\BETI}  contains one I$_3^-$ ion
at the (0 0 0) inversion site and two {\ET} molecules at generic
sites.\cite{Mori1984,Schultz1986,Leung1985} At 4.5 K the two {\ET}
molecules have a boat
geometry (with the terminal C atoms in 9a,10a positions) and are
interconverted by the inversion.\cite{Schultz1986}  At 100 and 120 K the
lattice is disordered\cite{Leung1985} and inversion symmetry is satisfied only
statistically, with a mixture of boat and chair molecules.

As explained in section II, at first we have made calculations with rigid
molecules, and then we have relaxed the RMA with the addition of a subset of
{\it intra}--molecular modes. The crystal structure is only marginally
affected by RMA and in Table II we report the comparison between
RMA-calculated and experimental crystal
structure\cite{Schultz1986,Leung1985} at
several temperatures and pressures. Fig. 1 reports a more extensive
and direct comparison between calculated and experimental crystal
axis lengths against $T$ and $p$.
The calculations have been performed by minimizing the
free energy $G$ with the molecules kept rigid at their experimental, ordered
geometry at 4.5 K.\cite{Schultz1986} To investigate the effect of small changes in
molecular geometry, the structures of {\BETI} have been recomputed with the
120 K geometry\cite{Leung1985} and ordered molecules (staggered or boat form).
The effect of the change in molecular geometry is negligible. At all
temperatures, {\BETI} appears to be thermodynamically more stable
than {\AETI},\cite{Valle1999} giving account for the irreversible
interconversions of {\AETI} into $\beta$--like phases.

The effect of molecular deformations has been investigated,
within the RMA approximation,
by testing several model geometries
in {\BETI} as well as in {\AETI} phases.
The potential energy has been minimized with the experimental
geometries\cite{Leung1985,Endres1986}
and with chair-$\alpha$, chair-$\beta$, boat-$\alpha$ and
boat-$\beta$ model geometries. The chair-$\alpha$ geometry is the average
of the two chair molecules in {\AETI},\cite{Endres1986} while chair-$\beta$ is
the molecule of {\BETI} with all the terminal carbons in 9b,10b
positions.\cite{Leung1985} The boat-$\alpha$ and $\beta$
geometries are those observed in
the corresponding phases.\cite{Leung1985,Endres1986}
For both $\alpha$ and $\beta$ phases, the potential
energy minimum is found with experimental geometry of
that phase, and the system becomes less stable if any other
geometry is used. It should be noticed that for {\BETI} the
boat-$\beta$ geometry coincides with the
experimental geometry at 4.5 K, and thus yields
the lowest energy. The difference between {\AETI} and {\BETI}
essentially vanishes if the chair-$\alpha$ and boat-$\alpha$ geometries
are used in the $\beta$--phase,
while the chair-$\beta$
and boat-$\beta$ geometries drastically destabilize the $\alpha$--phase.
This behavior clearly indicates that
molecular deformations play a crucial role in stabilizing the
various {\ETI} phases.

\subsection{Specific heat}

We next turn our attention to the phonon structure.
In the $\beta$--phase we have only one formula unit in the triclinic unit
cell, and within RMA we expect 8$A_g$ and 6$A_u$ $\q = 0$ lattice
phonons active in Raman and in IR, respectively. The number of
phonons experimentally observed in the
10-150 cm$^{-1}$ spectral region is in any case smaller than the
above prediction, so vibrational spectra do not offer a very
stringent test of the calculations. On the other hand, there is another
observable, the specific heat, which depends on the frequency
distribution. As shown in Fig. 2, at 20 K the $C_V$ calculated within RMA
(dotted line) is about 50\% smaller than the experimental $C_p$
(dots, from Ref.\onlinecite{Stewart1986}).
The difference between $C_V$ and $C_p$ is usually
small for solids, since their thermal expansion is small. Therefore,
we attribute most of the discrepancy between $C_V$ and $C_p$ to the
{\it intra}--molecular modes, which are neglected in the RMA calculation.

{\it Ab--initio} calculations \cite{Demiralp1997,Liu1997} indeed indicate the
presence of several low-frequency {\ET} {\it intra}--molecular
(internal) normal modes. Since calculations refer to a free molecule, a direct
comparison with experimental data in the solid state is not feasible.
However, they constitute a very convenient starting point for relaxing
RMA in QHLD calculations, as explained in Section II. We have
included the lowest nine {\ET} internal modes which fall in the same
spectral region as the lattice modes (below $\sim$ 220
cm$^{-1}$),\cite{Liu1997} and therefore are likely
coupled. In addition, the symmetric and antisymmetric
stretchings, and the two bendings of I$_3^-$, expected at 114, 145, 52
and 52 cm$^{-1}$, respectively,\cite{Lyndenbell1998} have been included
in the QHLD calculations. The cartesian displacements of {\ET}
were obtained from the {\it ab--initio} calculations,\cite{Liu} while those of
I$_3^-$ were determined by symmetry alone, as often it happens for
small molecules with high symmetry.

The $C_V$ computed by relaxing the RMA is also shown in Fig. 2. The agreement
with experiment is greatly improved with respect to RMA calculations. We
anticipate that the same kind of result has been obtained for
{\KETI},\cite{Girlando} and conclude by stating that RMA has to be
relaxed for a realistic calculation of the low-frequency phonons of
{\ET} crystals.

\subsection{Phonon assignments}

We now go back to the characterization of individual low-frequency
phonons. In the RMA classification, below 220 cm$^{-1}$ we expect
Raman activity for 8 lattice
modes, 9 {\ET} {\it intra}--molecular modes and one stretching of I$_3^-$;
in IR we expect 6 lattice modes, 9 {\ET} {\it intra}--molecular modes
and three I$_3^-$ modes. The modes calculated at the minimum $G$
structure at 120 K are compared with experimental
ones,\cite{Pokhodnia1993,Dressel1992,Swietlik1987} in Tables III
and IV for $A_u$ and $A_g$ modes, respectively.
We have chosen the 120 K temperature since in this way we can
compare the normal state phonon frequencies and eigenvectors for the
minimum $G$ and the experimental\cite{Leung1985} structure. The frequency
differences between minimum $G$ and experimental structure are quite small.
The comparison between calculated and
experimental vibrational frequencies is
satisfactory, although not very significant given the low number of
observed frequencies. Since we also have all the
corresponding eigenvectors, we report an approximate description of
the phonons, given for both {\ET} and I$_3^-$ as percentage of the
lattice (rigid molecule) and of the {\it intra}--molecular contributions.
Since in some cases there is a considerable mixing
between lattice and molecular modes, a clear distinction
cannot be made. Fig. 3 reports the full dispersion curves
along the C, V, X, and Y directions\cite{Mori1984} and
density of states of {\BETI}. In order to make the figure more
readable, we have limited the highest frequency to 150 cm$^{-1}$.
Fig. 3 puts in evidence the complex structure of {\BETI} low-frequency
phonons. We have a very dense grouping of modes in the 50-100 cm$^{-1}$
region, with several avoided crossings between the dispersion curves, and
clear mixing between lattice and molecular modes. Only the
acoustic phonon branches contribute to the density of states below
$\sim$ 25 cm$^{-1}$, so that the typical $\omega^2$ dependence
is observed. On the other hand, at energies higher than $\sim $
140 cm$^{-1}$ the almost dispersionless {\it intra}--molecular modes
dominate, and the phonon density of states
appears as a sum of delta-like peaks (not shown in the figure).

\subsection{Electron-phonon coupling}

The {\it e--lph} coupling constants for the optical
phonons of {\BETI} are reported in Table IV.
As explained in Section II, if one assumes that the
eigenvectors are independent of $\q$, only the
$A_g$ phonons can couple to electrons. In Table IV for each phonon
we report both the individual $g(KL;j,\q)$ (Eq. 5) and the
small polaron binding energy $\epsilon_j$.
The two most strongly coupled modes
are those calculated at 32 and 113 cm$^{-1}$. Whereas
the former mode has been observed in the Raman spectrum,
and together with the lower frequency mode (27 cm$^{-1}$)
undergoes a drastic intensity weakening at $T_c$,\cite{Pokhodnia1993}
the latter has not been reported even in the normal
state.\cite{Pokhodnia1993,Swietlik1987} The reason might be due to the
proximity of the very intense, resonantly enhanced band at 121
cm$^{-1}$, due to the symmetric stretch of the I$_3^-$ anion.
One band at 107 cm$^{-1}$ has been observed below 6 K for
488 nm laser excitation.\cite{Swietlik} On the other hand, a band at
109 cm$^{-1}$, whose intensity varies with sample and irradiation,
has been attributed to the splitting of the
I$_3^-$ stretching mode,\cite{Swietlik1987} as a consequence
of the commensurate superstructure reported in one X-ray
investigation at 100K.\cite{Endres1986} Certainly the 100-130 cm$^{-1}$
spectral region deserves further experimental scrutiny with the latest
generation of Raman spectrometers. A second observation is that whereas
the 113 cm$^{-1}$ mode involves only the {\ET} units, and is mostly a
lattice mode, the 32 cm$^{-1}$ one is a mixing between rigid I$_3^-$
motion and ``flexible'' {\ET} vibrations. This finding
suggests a not marginal role of the counter-ions sheets
in $\beta $--type {\ET} salts.

As shown by Table IV, the coupling of individual optical modes
with electrons is in general not particularly strong, but on the
whole the strength of {\it e--lph} coupling, as measured by the
sum of the $\epsilon_j$, is appreciable, around 45 meV,
to which we have to add the contribution of the acoustic phonons.
For the sake of comparison, we give also the
$\epsilon_j^{ac}$ for the three acoustic branches,
calculated as average over several points at the BZ edges.
In order of decreasing phonon frequency (see Fig. 3)
the $\epsilon^{ac}$ are: 2.3, 3.3 and 18.2 meV, respectively.
The coupling strength of the lowest acoustic branch
at the zone edge is comparable to that of the most
strongly coupled optical phonons.
Thus the overall {\it e-lph} coupling strength is of the same order
of magnitude as that due to {\it e--mv} coupling,
about 70 meV.\cite{Visentini1998}

We can make a more direct connection with superconducting
properties by calculating the Eliashberg function and the
dimensionless electron-phonon coupling constant $\lambda$.
As seen in eq. (6), the absolute value of the Eliashberg
function depends on the electronic density of states
at the Fermi energy, $N(E_F)$. Experimental estimates of this
critical parameter are problematic, since the measured
quantities already include or the $\lambda$ enhancement factor,
or the Coulomb enhancement factor, or both.
The available theoretical estimates are all based on the extended H\"uckel
tight binding method. The choice of $N(E_F)$ = 2.1
spin states/eV/unit cell, as obtained by this method,\cite{Haddon1994}
is consisten with our extended H\"uckel estimates of the CT integrals in real space.
To our advantage, we can compare the calculated
$\alpha(\omega)F(\omega)$ with that derived from normal state
current/voltage measurements at a point contact
junction.\cite{Nowack1986,Nowack} This kind of experiment is rather
difficult to perform on organic crystals like {\BETI},
since one has to be careful about pressure effects
at the point contact. The use of a contact between two {\BETI}
crystals made the current-voltage characteristics rather stable,
from which the $\alpha(\omega)F(\omega)$ function reported
in the upper part of Fig. 4 was obtained.\cite{Nowack}
We have changed the scale on the ordinate axis to maintain
the same energy unit (cm$^{-1}$) throughout.
It is clear that the spectral resolution of the experiment
is larger than  $\sim$ 10 cm$^{-1}$, and probably increases with energy.
Indeed, no spectral detail is visible beyond 240 cm$^{-1}$,
where the contribution of {\it e--mv} coupled modes should
be detectable.\cite{Ishiguro} Therefore, to make easier
the visual comparison with the experimental data, we have
smoothed the calculated  $\alpha(\omega)F(\omega)$
(Fig. 4, lower part) by a convolution  with a
Gaussian distribution. We have also assumed that the Gaussian
distribution width increases linearly with $\omega$ (from
0.1 to 20 cm$^{-1}$ in the 1-200 cm$^{-1}$ interval).

Fig. 4 puts in evidence the very good agreement between experiment
and calculation. The absolute scale of $\alpha(\omega)F(\omega)$
turns out to be practically the same, even if both experiment
and calculation  are affected by considerable uncertainties,
as explained above.
The three main peaks observed in the experiment are well
reproduced and are identified as due to the most strongly
coupled phonon branches, namely, the optical phonons at 113 and
32 cm$^{-1}$, and the lowest frequency acoustic branch.
The calculated peak frequency due to the latter 
is slightly higher than the experimental one (22 vs 10 cm$^{-1}$).
This discrepancy might be due to the fact that the experiment
refers to the $\beta_L$-{\ETI} phase, whereas our calculation refers
to a perfectly ordered phase like the $\beta^*$-{\ETI}.\cite{Nowack}
We also remark that, at variance with traditional superconductors,
the Eliashberg function is remarkably different from the phonon
density of states (Fig. 3). For instance, the peak around 120
cm$^{-1}$ in $F(\omega)$ is due to the dispersionless I$_3^-$
stretching mode, which is completely decoupled from the
electron system, whereas the broad peak in $\alpha(\omega)F(\omega)$
is due to the nearby (113 cm$^{-1}$ ``lattice'' mode of the {\ET} molecules.
Due to the complex phonon structure, $\alpha(\omega)$ is
not nearly constant, but varies rapidly with the frequency.

The dimensionless coupling constant $\lambda$ obtained by integration
of $\alpha(\omega)F(\omega)/\omega$ up to 240 cm$^{-1}$
turns out to be around 0.4. The contribution to
$\lambda$ from {\it e--mv} coupled modes\cite{Visentini1998}
is instead around 0.1. Thus in the McMillan picture the
overall $\lambda$ of {\BETI} \cite{Pedron1993}
is $\sim$ 0.5, which may well account
for the observed $T_c = 8.1$ K.\cite{Nowack}

\section{Discussion and conclusions}

The computational methods we have adopted to analyze the
crystal and lattice phonon structure, and the electron-phonon
coupling strength of {\BETI} are empirical or semiempirical.
The form of QHLD atom--atom potentials has no rigorous
theoretical justification, and the corresponding parameters
are derived from empirical fittings. We have adopted {\it ab--initio}
atomic charges\cite{Demiralp1994} to take into account Coulomb interactions
between atoms, and {\it ab--initio} vibrational eigenvectors\cite{Liu1997}
to introduce the coupling between lattice and molecular modes.
Also the extended H\"uckel method used to characterize the
{\BETI} electronic structure is semiempirical, albeit
with an experimental basis far wider than QHLD.
In view of the obvious limitations of empirical or
semiempirical methods, the success achieved in the case
of {\BETI} is even beyond our expectations, also
considering that {\it none} of the empirical parameters
has been adjusted to fit {\BETI} experimental data.

Indeed, {\it all} the available {\BETI} experimental data
have been accounted for. The crystal structure, and its
variation with temperature and pressure, is correctly reproduced
(Fig. 1 and Table II). Useful hints about the relative
thermodynamic stability of {\ETI} $\alpha$ and $\beta$
phases have been obtained, as well as some indications
on the effect of {\ET} conformation of the phases stability.
The specific heat (Fig. 2) and the few detected Raman
and infrared bands have been accounted for by including the
coupling with low-frequency molecular vibrations.
Finally, the point contact Eliashberg spectral function
has been satisfactorily reproduced (Fig. 4).

Despite the success, it is wise to keep in mind the QHLD
limitations. First and foremost, conformational disorder
in the crystal structure is not included. This is not a
limitation of the QHLD method only, but it is a serious one
particularly for {\BETI} salts, where disorder plays
an important role even in the superconducting properties.
It is in fact believed that a fully ordered structure is
at the origin of the higher $T_c$ (8.1 K) displayed by $\beta^*$-
or $\beta_H$- phases with respect to $\beta_L${\ETI}
(1.5 K).\cite{Schultz1986}
Furthermore, even if the QHLD method is able to follow
the $T$ and $p$ dependence of the crystal structure,
phase transitions implying subtle structural changes
may be beyond its present capabilities, even for fully ordered
structures. The relative stabilities of the phases can indeed
be reproduced only at a qualitative level. For what concerns
the electron-phonon coupling, one has to keep in mind that it
depends on phonon eigenvectors, and these are obviously
more prone to inaccuracies than the energies.
Finally, extension to other {\ET} salts with counter-ions different
from I$_3^-$ is not obvious, requiring additional atom-atom parameters.

Once the above necessary words of caution about the method are
spelled out, we can underline what in any case have learned
from the present QHLD calculations. So far, in the lack of
any description, no matter how approximate, of the phonons
modulating the CT integrals, only speculative discussions
about their role in the superconductivity could be put forward,
catching at best only part of the correct picture. One of the
most important indications coming out from the present paper
is the need of relaxing the RMA. So from one side
we cannot try to focus on the isolated molecule {\it intra}--molecular
vibrations presumably modulating the CT
integral,\cite{Demiralp1995} and on the
other side that the ``librations'' of the rigid molecules \cite{Nowack1986}
lack of precise meaning. In other words, there is no simple or
intuitive picture of the phonons modulating the CT integrals.
Our results, and the overall mode mixing, suggest that also the
counter-ions vibrations may play a perhaps indirect role in the coupling.

The results of the present paper definitely assess the very important
role played by the low-frequency phonons in the superconducting
properties of {\ET} salts. {\it Both} acoustic and optic modes
modulating the CT integral are involved. The overall dimensionless
coupling constant is $\sim 0.4$, much larger then that due
to {it e--mv} coupled phonons ($\sim 0.1$). Of course, a mere numerical
comparison of the two $\lambda$'s is not particularly significant,
since one has to keep in mind the very different time scales
(frequencies) of the two types of phonons. The phonons appreciably
modulating the CT integrals fall in the 0-120 cm$^{-1}$ spectral region
(Table IV), whereas those modulating on-site energies have
frequencies ranging from 400 to 1500 cm$^{-1}$.\cite{Visentini1998}
Applicability of the Migdal theorem to the latter appears dubious:
non-adiabatic corrections\cite{Pietronero1992} or alternative mechanisms
such as polaron narrowing\cite{Feinberg1990} have been
suggested. For these reasons we will not get involved into
detailed discussions about the relative role of {\it e--lph}
and {\it e--mv} coupling in the superconductivity mechanism.
We limit ourselves to state that phonon mediated coupling can
well account for the observed critical temperature of the
ordered {\BETI} phase, given plausible values of the other
fundamental parameter, the Coulomb pseudopotential
$\mu$.\cite{Pedron1993,Nowack}

The results of the present paper suggest that
phonon mediated mechanism is responsible for
the superconductivity of {\ET}-based salts. The same conclusion
was reached on the basis of the solution of the BCS gap equation
for $\kappa$ phase {\ET} salts.\cite{VisentiniK}
On the other hand, evidences are also accumulating towards non-conventional
coupling mechanisms in organic superconductors, such as
spin-fluctuation mediated superconductivity.\cite{Schrama1999} Similar
apparently contrasting experimental evidences are also
found for cuprates,\cite{Scalapino1995} pointing to a
superconductivity mechanism
where {\it both} electron-phonon coupling and antiferromagnetic spin
correlations are taken into account.

\acknowledgements
We express many thanks to  Rufieng Liu for
providing the {\it ab--initio} cartesian displacements of {\ET}, to
R. Swietlick for sending us unpublished Raman spectra of {\BETI},
and to A.M\"uller for useful correspondence. We acknowledge
helpful discussions with many people, notably A.Painelli, D.Pedron,
D.Schweitzer, H.H.Wang, J.Wosnitza.
This work has been supported by the Italian National
Reasearch Council (CNR) within its ``Progetto Finalizzato
Materiali Speciali per tecnologie Avanzate II'', and the
Ministry of University and of Scientific and Technological
Research (MURST).

\eject
\mediumtext
\begin{table}[ht]
\caption{Parameters for the atom-atom potential $ V_{ij}(r) = A_{ij}\exp(-B_{ij}
r) - C_{ij}/r^6 $. $V$ is in kcal/mol, $r$ in \AA, $A$, $B$, $C$ in
consistent units. Heteroatom parameters are given by
$A_{ij}=\protect\sqrt{A_{ii}A_{jj}}$, $B_{ij}=(B_{ii}+B_{jj})/2$ and
$C_{ij}=\protect\sqrt{C_{ii}C_{jj}}$.}
\narrowtext
\begin{tabular}{lrrr}
$i$ & $A_{ii}$~ & ~$B_{ii}$~ & $C_{ii}$~ \\
\hline
H   &      2868 & 3.74 &    40.2 \\
C   &     71460 & 3.60 &   449.3 \\
S   &    329600 & 3.31 &  5392.0 \\
I   &    642376 & 3.09 & 16482.6 \\
\end{tabular}
\widetext
\end{table}

\begin{table}[ht]
\caption{Structural data of {\BETI} as a function of $T$ (K) and $p$
(GPa): experimental (Ref. \protect\onlinecite{Schultz1986}
and computed unit cell axis
$a$, $b$, $c$ (\AA), angles $\alpha$, $\beta$, $\gamma$ (degrees) and
volume $V$ (\AA$^3$). The lattice is triclinic, space group
$P\overline{1}$ ($C_i^1$), with $Z=1$.}
\begin{tabular}{cclcccccccc}
$T$   & $p$ &       & $a$   & $b$ & $c$ & $\alpha$ & $\beta$ & $\gamma$ & $V$ & G(p,T) \\
\hline
4.5   & 0   & expt. & 6.519 & 8.920 & 15.052 & 95.32 & 96.09 & 110.44 & 807.6 & \\
      &     & calc. & 6.571 & 9.147 & 15.073 & 93.94 & 95.07 & 111.85 & 832.6 & $-229.034$ \\
20    & 0   & expt. & 6.543 & 8.968 & 15.114 & 95.34 & 96.05 & 110.30 & 819.1 & \\
      &     & calc. & 6.571 & 9.147 & 15.074 & 93.94 & 95.07 & 111.85 & 832.6 & $-229.056$ \\
120   & 0   & expt. & 6.561 & 9.013 & 15.173 & 95.07 & 95.93 & 110.28 & 829.2 & \\
      &     & calc. & 6.577 & 9.159 & 15.095 & 93.85 & 95.07 & 111.82 & 835.9 & $-231.677$ \\
298   & 0   & expt. & 6.615 & 9.100 & 15.286 & 94.38 & 95.59 & 109.78 & 855.9 & \\
      &     & calc. & 6.591 & 9.189 & 15.145 & 93.64 & 95.08 & 111.72 & 844.2 & $-242.678$ \\
4.5   & 1.5 & expt. & 6.449 & 8.986 & 15.034 & 94.79 & 96.57 & 111.29 & 799.1 & \\
      &     & calc. & 6.556 & 9.117 & 15.048 & 94.05 & 95.14 & 111.85 & 826.1 & $-211.930$ \\
6.1   & 4.6 & expt. & 6.433 & 8.947 & 14.927 & 95.15 & 96.77 & 111.40 & 786.1 & \\
      &     & calc. & 6.532 & 9.069 & 15.007 & 94.20 & 95.27 & 111.86 & 816.0 & $-175.319$ \\
\end{tabular}
\end{table}

\begin{table}
\caption{Low energy $A_u$ phonons
of $\beta$--(\ET)$_2^+$I$_3^-$.}
\begin{tabular}{dddddd}
\multicolumn{1}{c}{Expt.\tablenote{From Ref.\onlinecite{Dressel1992}}}
& \multicolumn{1}{c}{Calc.} &
\multicolumn{4}{c}{Approximate description} \\
\cline{1-2} \cline{3-6}
cm$^{-1}$ & cm$^{-1}$  &
\multicolumn{2}{c}{I$_3^-$ (\%)}&
\multicolumn{2}{c}{\ET$_2^+$ (\%)}\\
\cline{3-4} \cline{5-6}
 & & lattice & internal & lattice & internal \\
\tableline
  & 216 & & &  & 98 \\
  & 173 & & &  & 90 \\
  & 151 & & & 26 & 68 \\
133  & 148 & &86 &  & 10 \\
130  & 136 & &8 &  & 88 \\
124  & 123 & & &48  & 48 \\
  & 114 & & &8  & 86 \\
95  & 112 & & &24  & 72 \\
91  & 90 & & & 56 & 40 \\
  & 84 &8 &13 & 34 & 44\\
  & 74 & &56 & 26 & 12 \\
71  & 69 & & &22  & 78 \\
  & 67 &8 &40 & 18 & 36 \\
  & 62 &22 &44 & 10 & 20 \\
  & 57 & &23 & 44 & 28 \\
48  & 43 &55 &11 &26  & \\
  & 33 &44 & & 22 & 30 \\
  & 17 &41 & &22  & 36 \\
\end{tabular}
\end{table}

\begin{table}
\caption{Low energy $A_g$ phonons and coupling constants
of $\beta$--(\ET)$_2^+$I$_3^-$.}
\begin{tabular}{ddddddrrrrd}
\multicolumn{1}{c}{Expt.
\tablenote{From 
Ref.\onlinecite{Pokhodnia1993,Swietlik1987}.}} &
\multicolumn{1}{c}{Calc.} &
\multicolumn{4}{c}{Approximate description} &
\multicolumn{4}{c}{Coupling Constants (meV)
\tablenote{The modulated hopping
integrals between dimers are labeled according
to Ref.\onlinecite{Mori1984}; the calculated equilibrium values are:
$t_{AB}=0.22$ eV,
$t_{AC}=0.08$ eV,
$t_{AE}=0.10$ eV,
$t_{AH}=0.06$ eV.}} &
\multicolumn{1}{c}{$\epsilon_j$, meV} \\
\cline{1-2} \cline{3-6} \cline{7-10}
cm$^{-1}$ & cm$^{-1}$  &
\multicolumn{2}{c}{I$_3^-$ (\%)}&
\multicolumn{2}{c}{\ET$_2^+$ (\%)}&
$g(AB;j) $ &
$g(AC;j) $ &
$g(AE;j) $ &
$g(AH;j) $ & \\
\cline{3-4} \cline{5-6}
 & & lattice & internal & lattice & internal & & & & & \\
\tableline
214 & 207 &    &    &    & 90 &       & $-$1      &  $-$1     &           &    \\
179 & 172 &    &    &    & 92 &       & $-$1      &  $-$1     &           &    \\
    & 165 &    &    & 16 & 84 & 3     &          &  2        & $-$1      &    \\
149 & 144 &    &    & 16 & 80 & $-$1  & 2         &  1        & $-$2      &    \\
    & 131 &    &    & 12 & 86 & $-$6  &          &           & 1         &    \\
121 & 120 &    & 89 &    &    & $-$3  & 1     &  1    & $-$1 &    \\
    & 113 &    &    & 58 & 32 & $-$13 & 7     &  3    & $-$1 & 16 \\
    & 111 &    &    &    & 90 & $-$4  & 1     &  $-$1 & 3    &    \\
 91 & 87  & 10 &    & 50 & 36 & 2     &       &  3    & 4    &    \\
    & 81  &    &    & 48 & 48 & $-$1  & $-$1  &  $-$1 & $-$5 &    \\
    & 73  &    &    & 8  & 88 & 2     &       &  2    & 1    &    \\
    & 64  &    &    & 78 & 22 & $-$4  & 6     &  $-$3 & 3    & 9  \\
    & 60  & 33 &    & 18 & 50 & 1     & $-$3  &  $-$4 & 5    & 7  \\
 53 & 51  & 25 &    & 50 & 22 & 4     &       &  $-$6 & 3    & 10 \\
    & 49  & 20 &    & 66 & 14 & 4     & $-$3  &  $-$1 & 2    &    \\
 39 & 44  & 33 &    & 60 &    & $-$1  & $-$2  &  $-$2 & 2    &    \\
 32 & 32  & 56 &    & 16 & 24 & 6     & $-$2  &  $-$5 & 1    & 17 \\
 27 & 29  & 10 &    & 76 & 10 & $-$4  & 4     &  2    & 2    & 11 \\
\end{tabular}
\end{table}

\vskip 50mm
\protect\includegraphics [scale=0.5]{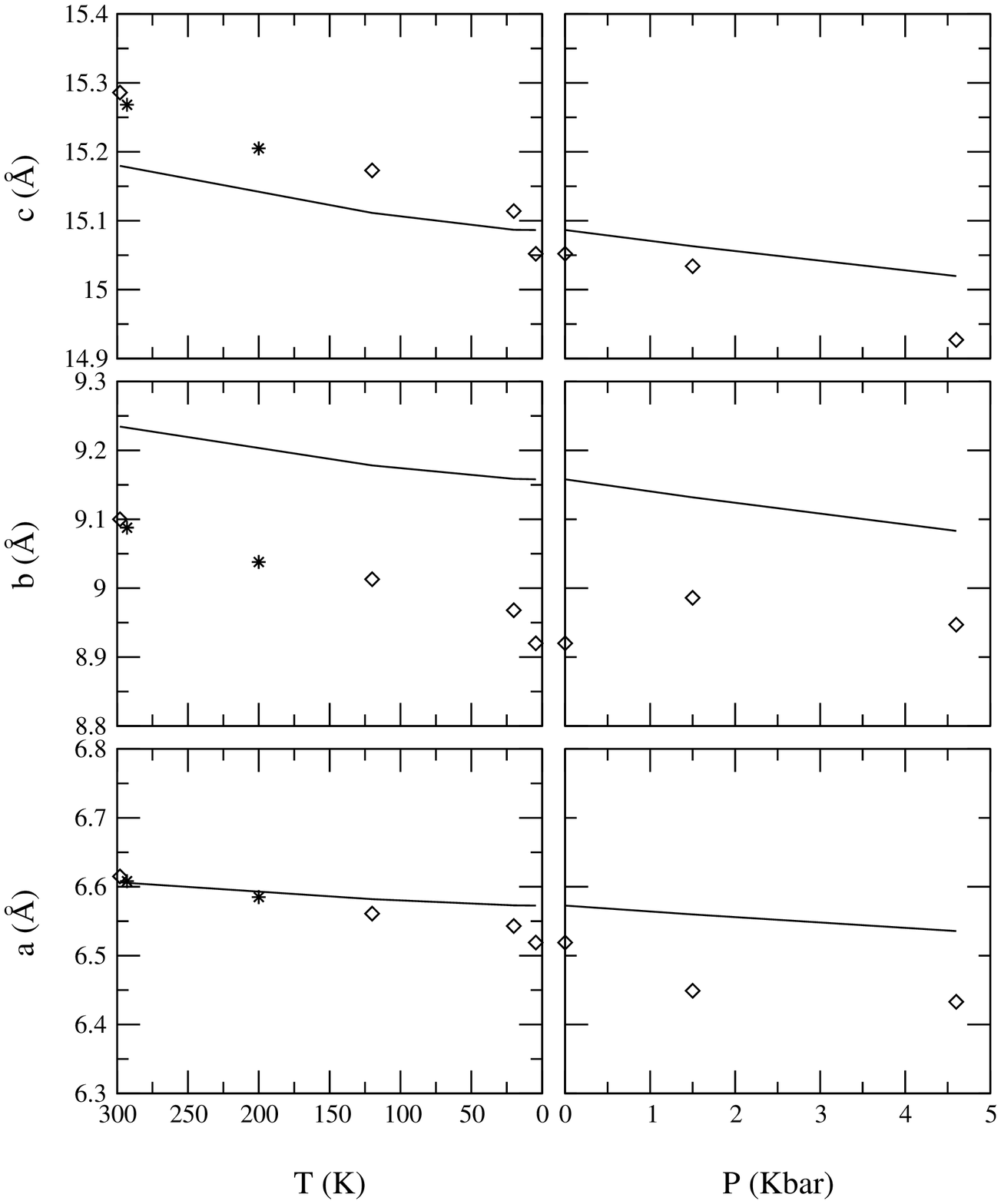}

\begin{figure}
\caption{Calculated and experimental of {\BETI} crystallographics axis
lengths as functions of temperature and pressure. The experimental
points are taken from Ref. \protect\onlinecite{Schultz1986} (diamonds)
and from Refs. \protect\onlinecite{Mueller1997,Mueller1999} (asterisks).}
\end{figure}

\newpage
\protect\includegraphics [scale=0.4,angle=-90]{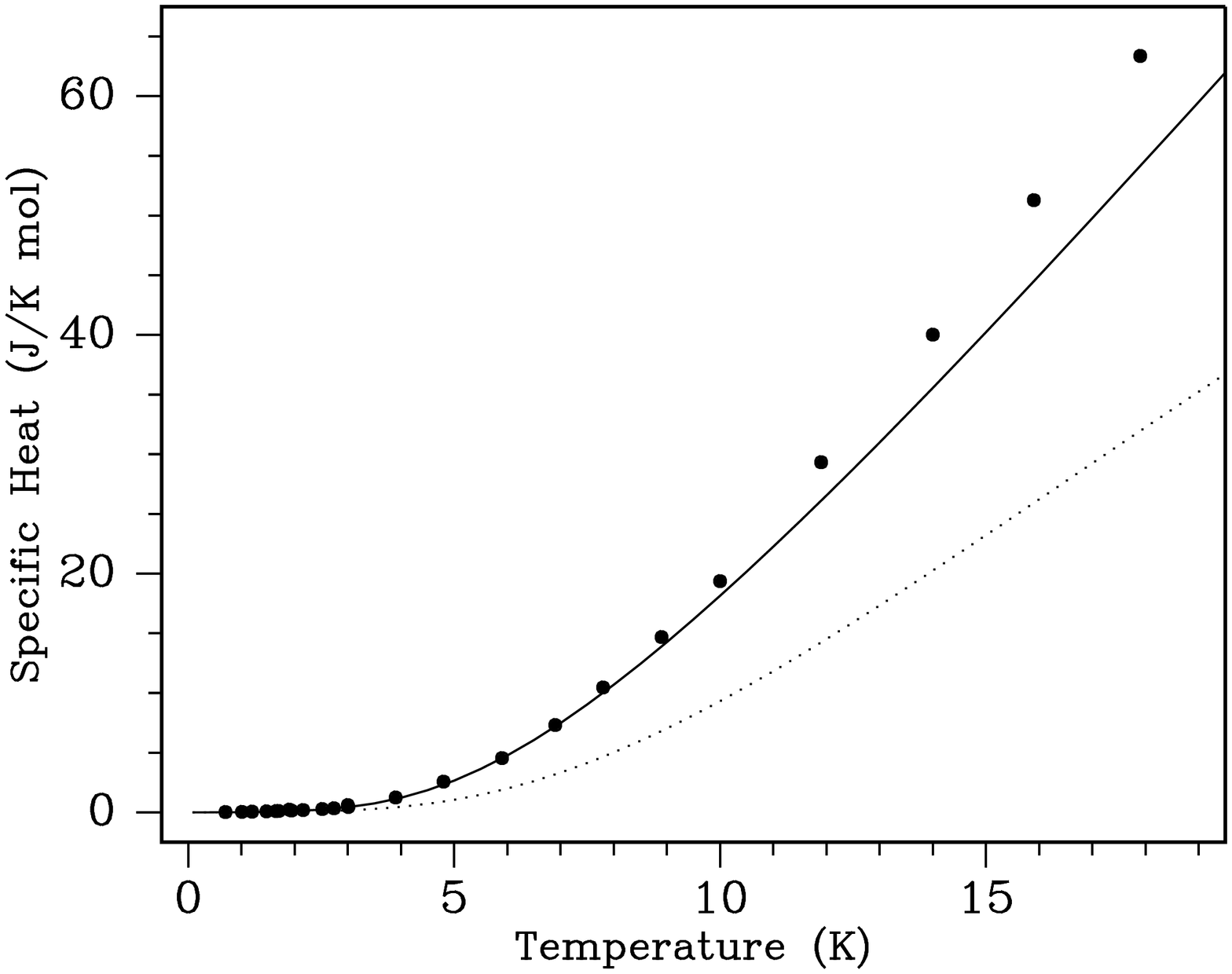}
\begin{figure}
\caption{Specific heat of {\BETI}
as a function of $T$.
The dots represent the experimental
$C_p$, from Ref. \protect\onlinecite{Stewart1986}.
The dotted line represents
computed $C_V$ due to lattice modes only in the RMA approximation, whereas
the dotted line is the $C_V$ obtained from coupled lattice and
{\it intra}-molecular modes.}
\end{figure}

\protect\includegraphics [scale=0.6,angle=-90]{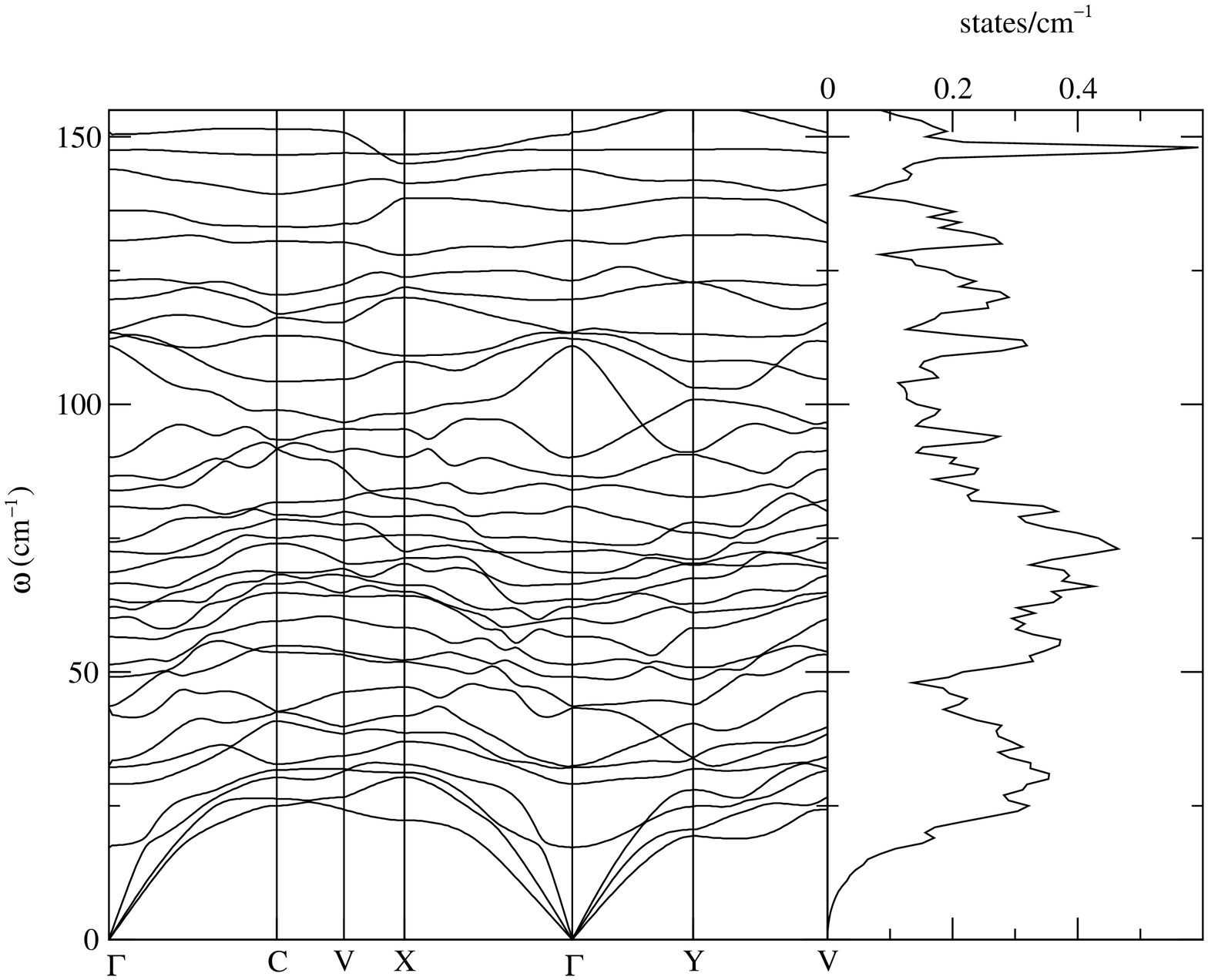}
\begin{figure}
\caption{Dispersion curves and density of states $F(\omega)$
of {\BETI} low-frequency phonons. The zone edges are labeled
according to Ref. \protect\onlinecite{Mori1984}.}
\end{figure}
\newpage
\protect\includegraphics [scale=0.5]{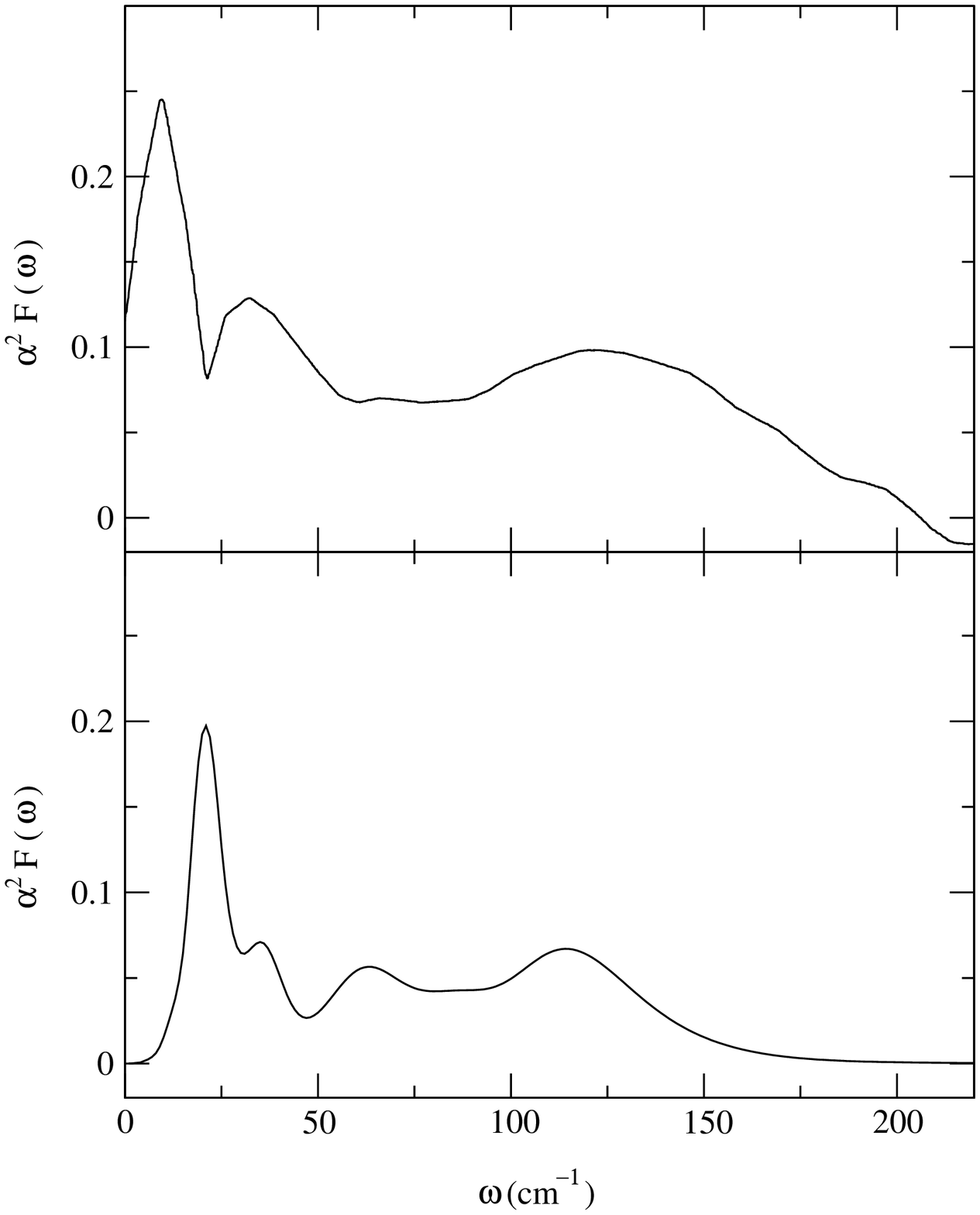}
\begin{figure}
\caption{Upper panel: the Eliashberg
function as measured from point--contact tunneling experiments
(adapted from Ref. \protect\onlinecite{Nowack}).
Lower panel: the calculated contribution
to $\alpha^2(\omega)F(\omega)$ from low frequency {\it e-lph} coupled phonons.}
\end{figure}

\end{document}